\documentclass[a4paper]{aa}
\usepackage[T1]{fontenc}
\usepackage[utf8]{inputenc}
\usepackage{graphicx}
\usepackage{color}
\usepackage{txfonts}
\usepackage{microtype}
\usepackage{ellipsis}
\usepackage{natbib}

\bibpunct{(}{)}{;}{a}{}{,}
\usepackage[pdftex,colorlinks=true,urlcolor=blue]{hyperref}
\usepackage[xindy,nonumberlist,nopostdot,nogroupskip]{glossaries}

\makeglossaries
\usepackage[xindy]{imakeidx}
\makeindex

\newcommand{\capurl}[1]{{\fontsize{8}{10}\selectfont\url{#1}}}

\begin{document}
\title{The Engine of Life}
\author{Markus Nielbock\inst{\ref{inst1}} \and Marco J.~Türk\inst{\ref{inst2},\ref{inst3}}}
\institute{Haus der Astronomie, Campus MPIA, Königstuhl 17, D-69117 
Heidelberg, Germany\\
\email{nielbock@hda-hd.de}\label{inst1}
\and
Universität Heidelberg, Fakultät für Physik und Astronomie, Im Neuenheimer Feld, D-69120 Heidelberg, Germany\label{inst2}
\and
Ottheinrich-Gymnasium, Gymnasiumstraße 1 -- 3, D-69168 Wiesloch, Germany\label{inst3}}

\date{Received February 18, 2016; accepted }

\abstract{This activity has been developed as a resource for the ``EU Space Awareness'' educational programme. As part of the suite ``Our Fragile Planet'' together with the ``Climate Box'' it addresses aspects of weather phenomena, the Earth's climate and climate change as well as Earth observation efforts like in the European ``Copernicus'' programme. This activity uses a simple analogue for the power of radiation received at a given distance from a star. A photovoltaic cell is connected to an electric motor. Depending on the power received on the cell, the motor begins to move. It changes also its speed with respect to the distance between the cell and the lamp. This can be interpreted as a model for a planetary system and its habitable zone. The ``engine of life'' moves as soon as the receiving power is big enough to sustain its operation. The distance, where the motor stops, can be interpreted as the outer edge of the habitable zone. As a second activity, the students will reconstruct the orbits of a real exoplanetary system by drawing a scaled model. In addition, they will calculate and superimpose a realistic circumstellar habitable zone and discuss its elements.}


\maketitle
%

\section{Background information}
\subsection{The Circumstellar Habitable Zone}
The most important ingredient to sustain life as we know it is liquid water. Therefore, if scientists want to find planets of other celestial bodies, where life may be present, they first want to know, if water exists in its liquid form there. This, for instance, is also one of the major goals when exploring and investigating Mars.

\begin{figure}
\centering
\resizebox{\hsize}{!}{\includegraphics{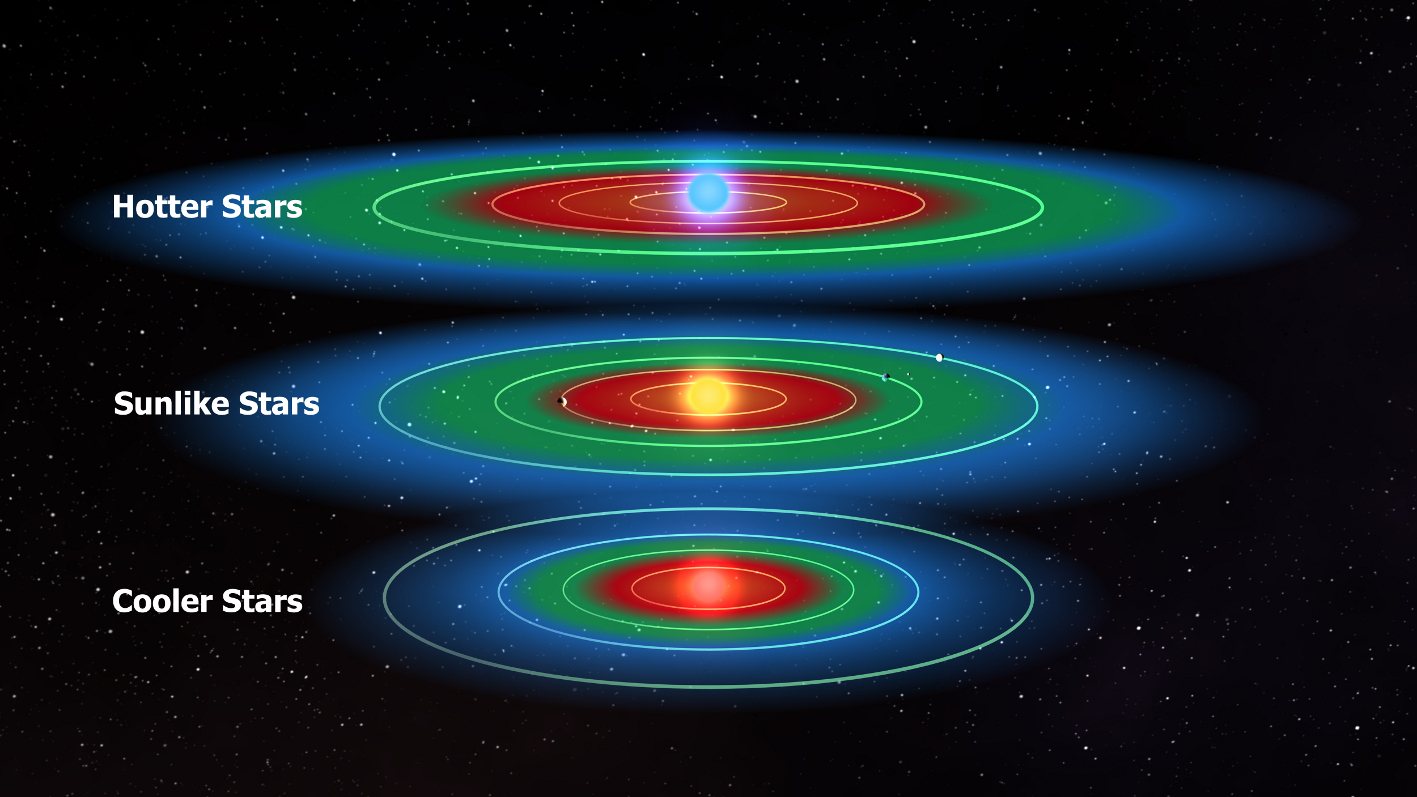}}
\caption{The circumstellar habitable zones of three stars that differ in size, luminosity and surface temperature
(NASA/Kepler Mission/D.~Berry, \capurl{http://aasnova.org/2016/02/24/where-to-look-for-habitability/}).}
\label{f:hz}
\end{figure}

The presence of liquid water depends on environmental conditions like air temperature and atmospheric pressure. The main driver of the surface temperatures of planets is their distance from the central star they orbit. Only in a small window the temperatures are just right so that water does not completely evaporate or freeze. These conditions are modified by local influences like the density of the atmosphere and the composition with potential greenhouse gases. This defines a range around a given star, in which liquid water could be present. This range is defined as the\newglossaryentry{hz}{name = {Habitable Zone}, description={In astronomy and astrobiology, the circumstellar habitable zone (CHZ), or simply the habitable zone, is the range of orbits around a star within which a planetary surface can support liquid water given sufficient atmospheric pressure.}}
{\em habitable zone} (Fig.~\ref{f:hz}). If a planet is found orbiting in this zone, it may potentially possess water in its liquid form and thus sustain life as we know it. In the solar system, only Earth and Mars are inside it.

\begin{figure}
\centering
\resizebox{\hsize}{!}{\includegraphics{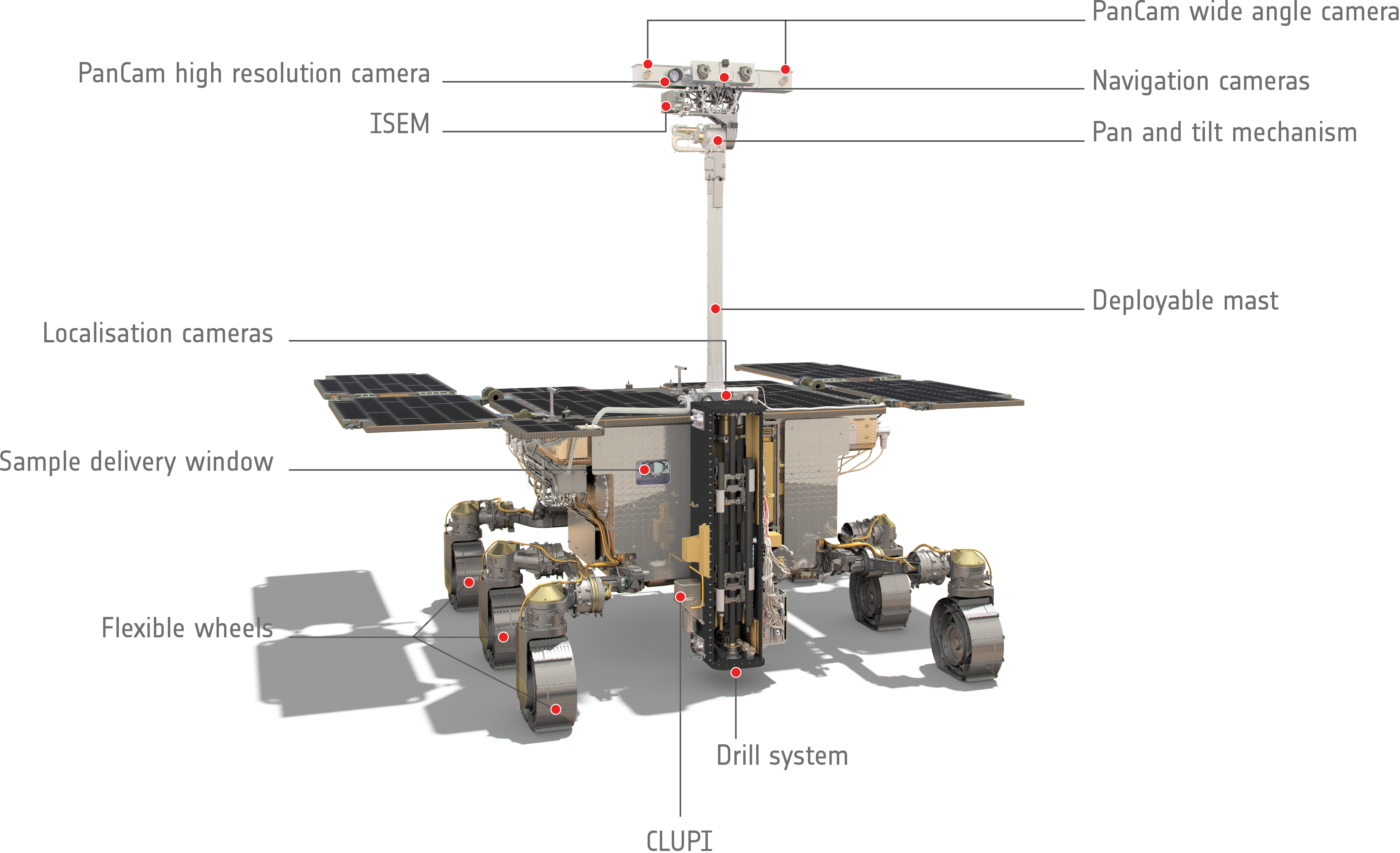}}
\caption{Computer model of the ExoMars rover to be launched in 2020 (ESA/ATG medialab, \capurl{http://exploration.esa.int/
mars/58885-exomars-rover-front-view-annotated/}, \capurl{http://www.esa.int/spaceinimages/ESA_Multimedia/Copyright\_Notice_Images}).}
\label{f:exomars}
\end{figure}

There is no guarantee that any planet orbiting inside the habitable zone actually possesses notable amounts of liquid water or harbours life, because the individual conditions for any given planet can be very different. Other boundary conditions that may help to sustain life are energy sources (light, chemical) and magnetic fields to fend off ionising particle radiation.

\subsection{Did or does Mars support life?}
Mars has half the size of the Earth. Its reddish colour is caused by iron oxide (rust) and has a very thin atmosphere, which mainly consists of carbon dioxide. One of its special features is its many extinct volcanoes, which reach heights of up to 22,000 metres! Like Earth, it also has seasons, as its axis of rotation is inclined. Theoretically, if the habitable zone of the solar system is considered, Mars has the potential to be conducive to life.

To date, there is no indication that Mars is or was inhabited. However, there is strong evidence that at an early age there has been much liquid water at the surface of that planet. The most common explanation for its current state is that Mars has lost its atmosphere so that most of the water either evaporated into space or is still present as ice deposits below the surface. Mars is so small that it exerts a rather low gravity that can barely hold the atmosphere. In addition, the lack of a magnetic
field made it easy for the solar wind to deplete the early Martian atmosphere.

Scientists hope that some form of life could have survived within the ice sheets below the Martian surface for some time. In particular, the European Mars programme {\em ExoMars} \citep[see Fig.~\ref{f:exomars}]{exomars} will investigate that hypothesis with a robotic laboratory that will be launched in 2020. It will be able to drill below the Martian surface and probe it for chemical and biological activity.

\begin{figure}
\centering
\resizebox{\hsize}{!}{\includegraphics{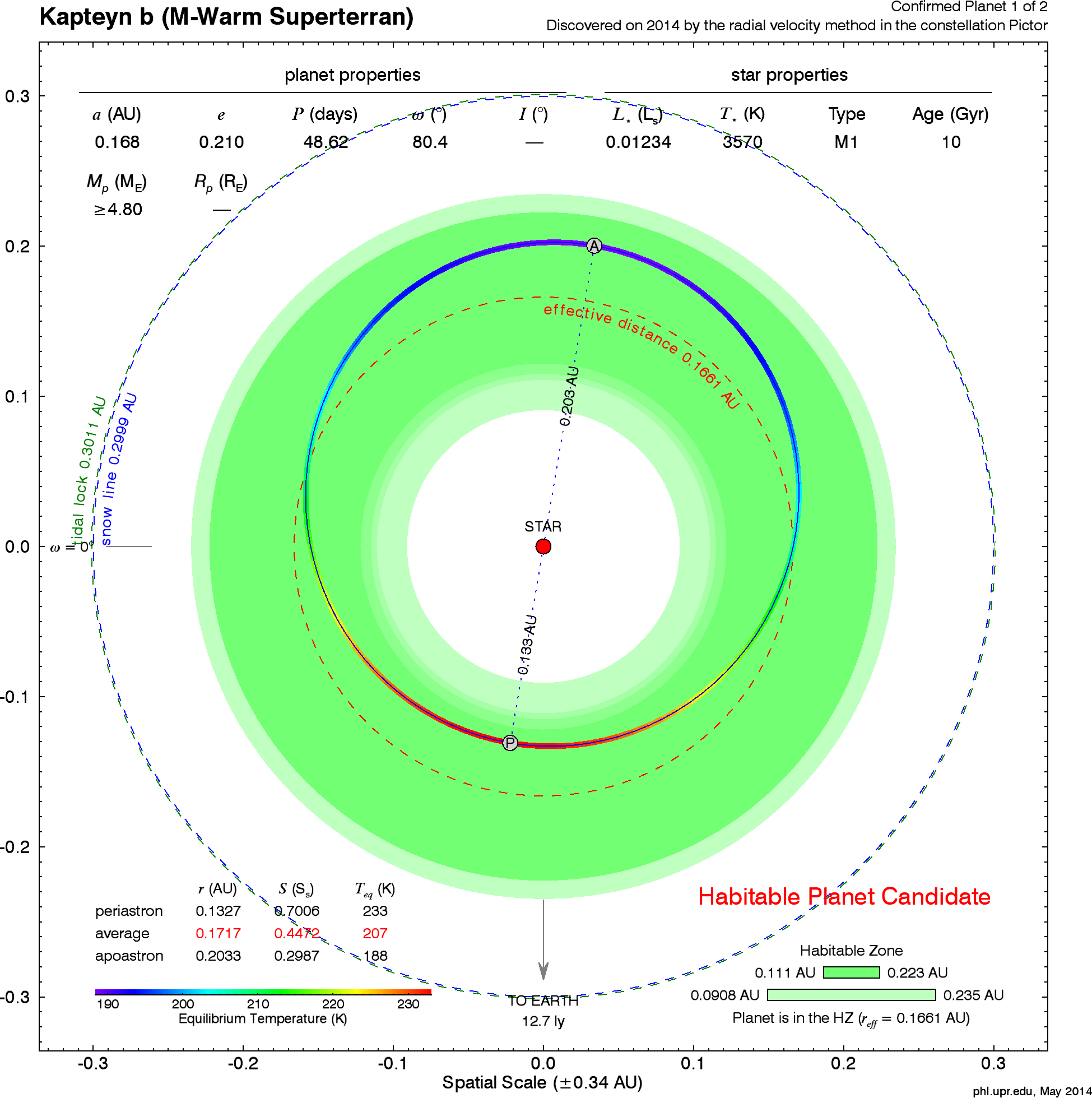}}
\caption{Habitable zone (green) around the star HD~33793. The orbit of the extrasolar planet Kapteyn~b is indicated (PHL/UCP Arecibo, \capurl{http://phl.upr.edu/press-releases/kapteyn}, \capurl{https://creativecommons.org/licenses/by-nc-sa/3.0/legalcode}, based on: \citet{2014MNRAS.443L..89A}).}
\label{f:kapB1}
\end{figure}

\subsection{Exoplanets}
Planets that orbit stars other than the Sun are called extrasolar planets or short\newglossaryentry{exoplanet}{name=Exoplanet,description={A planet that orbits a star different from the Sun.}} exoplanets. The first exoplanet hosted by a Sun-like star was discovered in 1995 by a Swiss team led by Michel Mayor and Didier Queloz from the University of Geneva \citep{1995Natur.378..355M}. This planet named 51 Pegasi b was anything but an Earth-like planet, as it revolves around its host star at a distance of only 0.05~AU, i.e.~5\% of the mean distance between the Sun and the Earth (1~AU = 1~Astronomical Unit).
\newglossaryentry{au}{name={Astronomical Unit (AU)},description={This is the mean distance between the Sun and the Earth. In astronomy, it is a convenient unit to describe distances within planetary systems.}} Note that even the planet closest to the Sun, Mercury, orbits it at a mean distance of 0.5 AU, i.e.~10 times farther away. In addition, it is a gas giant similar to Jupiter and Saturn. To date (July 2017), we know of 3627 exoplanets of which 2718 are planetary systems (see: \url{http://www.exoplanets.eu}).

Figure~\ref{f:hz} shows the circumstellar habitable zone (in green) for stars of different temperature and luminosity.
\newglossaryentry{L}{name=Luminosity,description={In astronomy, the total radiative power of a star is called {\em Luminosity}. It is measured in Watts.}} The star in the middle corresponds to stars similar to the Sun. Hotter stars usually have a larger and wider habitable zone, while cooler stars can only provide smaller and narrower habitable zones. The blue area indicates that it is too cold for liquid water and the red area indicates that it is too hot.

\begin{figure}
\centering
\resizebox{\hsize}{!}{\includegraphics{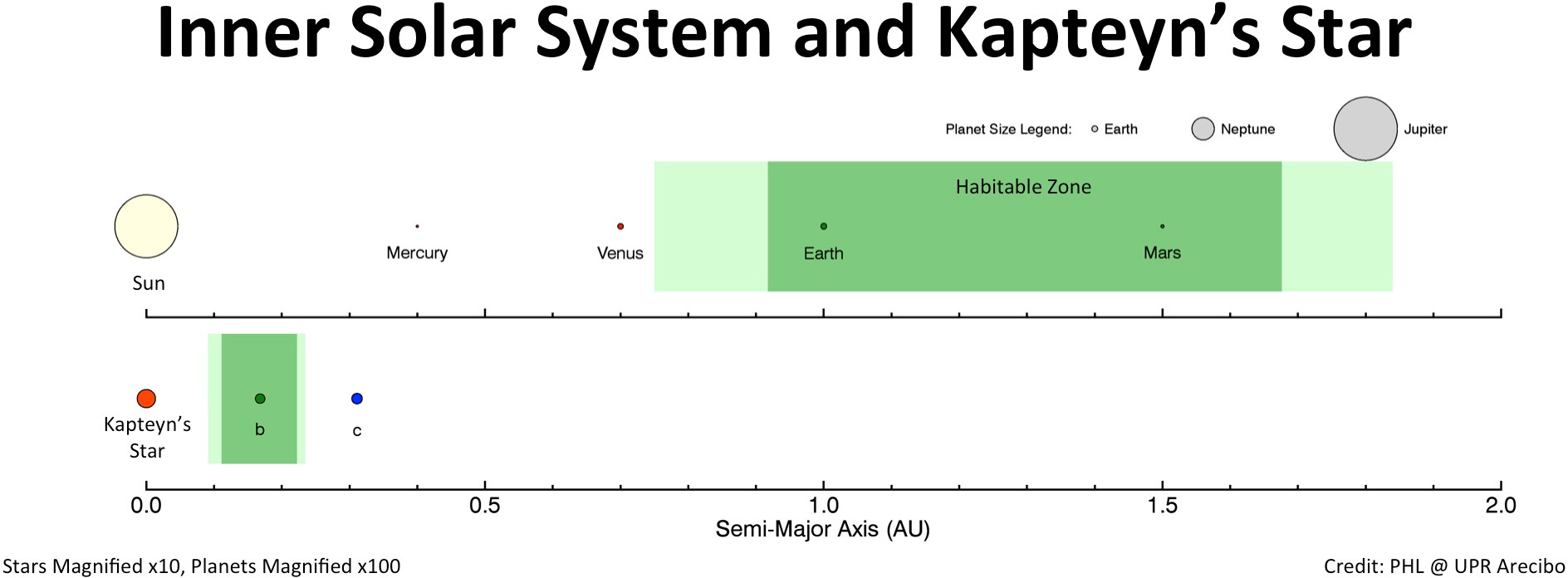}}
\caption{Comparison of the habitable zones (green) of the star HD~33793 and the Sun. (PHL/UCP Arecibo,
\capurl{http://phl.upr.edu/press-releases/kapteyn}, \capurl{https://creativecommons.org/licenses/by-nc-sa/3.0/legalcode}, based on: \citet{2014MNRAS.443L..89A}).}
\label{f:kapB2}
\end{figure}

Figure~\ref{f:kapB1} shows an example of an exoplanet around the star HD~33793, also known as Kapteyn’s star. The star has a surface temperature of about 3570~K. This is about half of the surface temperature of the Sun with 5778~K. Thus, the habitable zone (indicated in green) is located at a distance between 0.11~AU and 0.22~AU. Due to the lower luminosity, the habitable zone is closer to the star and narrower. When comparing the habitable zones of the Sun and HD~33793 (Fig.~\ref{f:kapB2}), we find that they both differ considerably. The solar habitable zone is both wider and farther away from the Sun.

The exoplanet Kapteyn~b with a mass of about 5 earth masses is located well within the habitable
zone and therefore is interesting for further research on habitability.

\subsection{Other locations in the Solar System}
During the recent years, scientists have been conveying the idea that there are other places in the solar system that might be habitable, although they are not within the habitable zone of the Sun. Especially icy moons such as Europa and Enceladus of the gas planets Jupiter and Saturn appear to be interesting in that respect. Observations made during solar system exploration missions like NASA’s {\em Cassini} (see: \url{https://saturn.jpl.nasa.gov/})have found evidence for subsurface oceans of liquid water below their thick ice shields \citep{redd_2015}. There is strong evidence that there are hydrothermal vents on the ocean floor of Enceladus. Similar findings have been made in the terrestrial deep sea. On Earth, these vents are colonised with life that feeds on hydrogen released from below \citep{ratner_2017}.

\begin{figure}
\centering
\resizebox{\hsize}{!}{\includegraphics{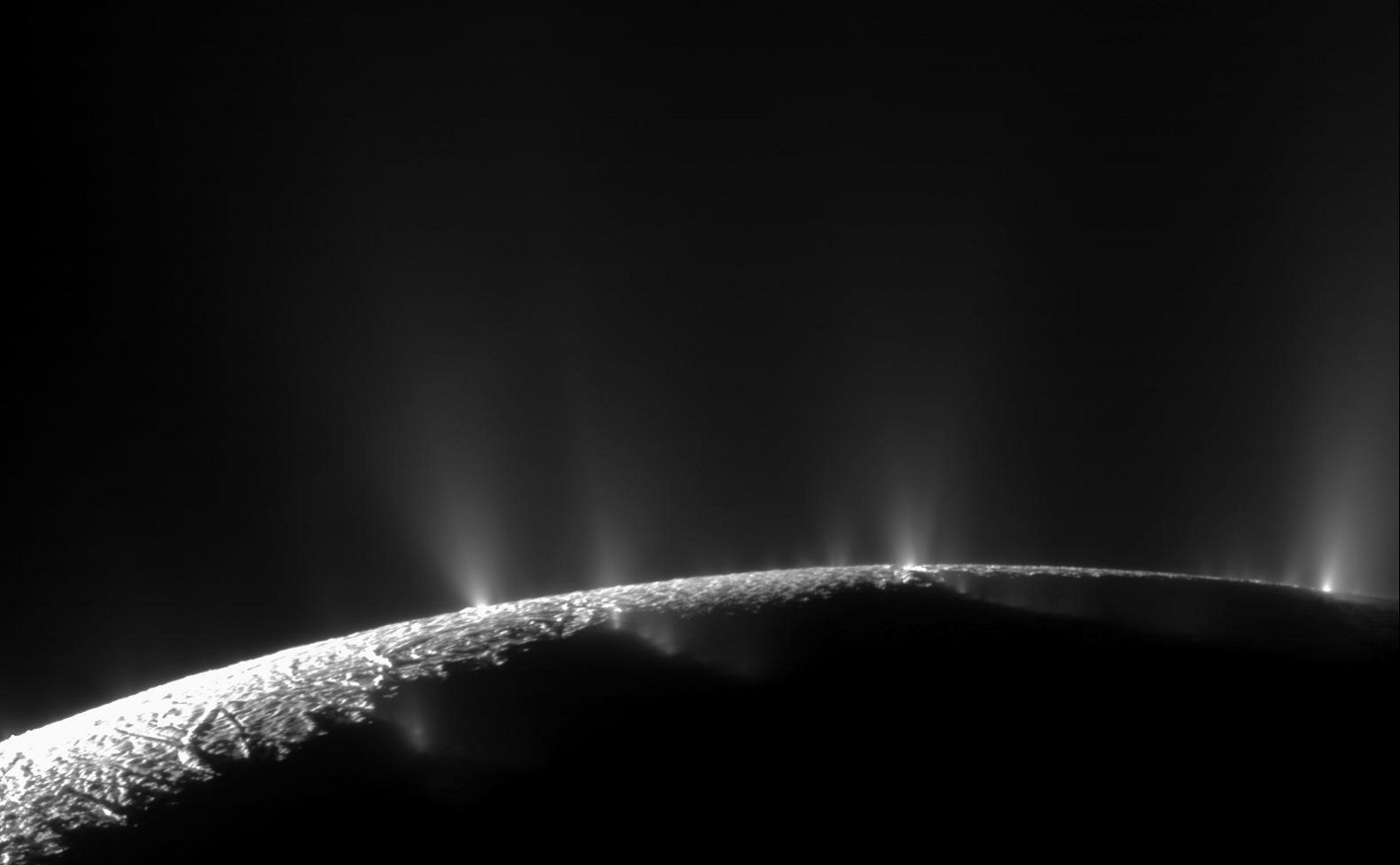}}
\caption{A dramatic plume sprays water ice and vapour from the south polar region of Saturn's moon Enceladus. Cassini's first hint of this plume came during the spacecraft's first close flyby of the icy moon on 17~February~2005 (NASA/JPL/Space Science Institute, \capurl{https://photojournal.jpl.nasa.gov/catalog/PIA11688}).}
\label{f:cryovulcano}
\end{figure}

\subsection{A simple formula for the habitable zone}
\begin{figure}
\centering
\resizebox{\hsize}{!}{\includegraphics{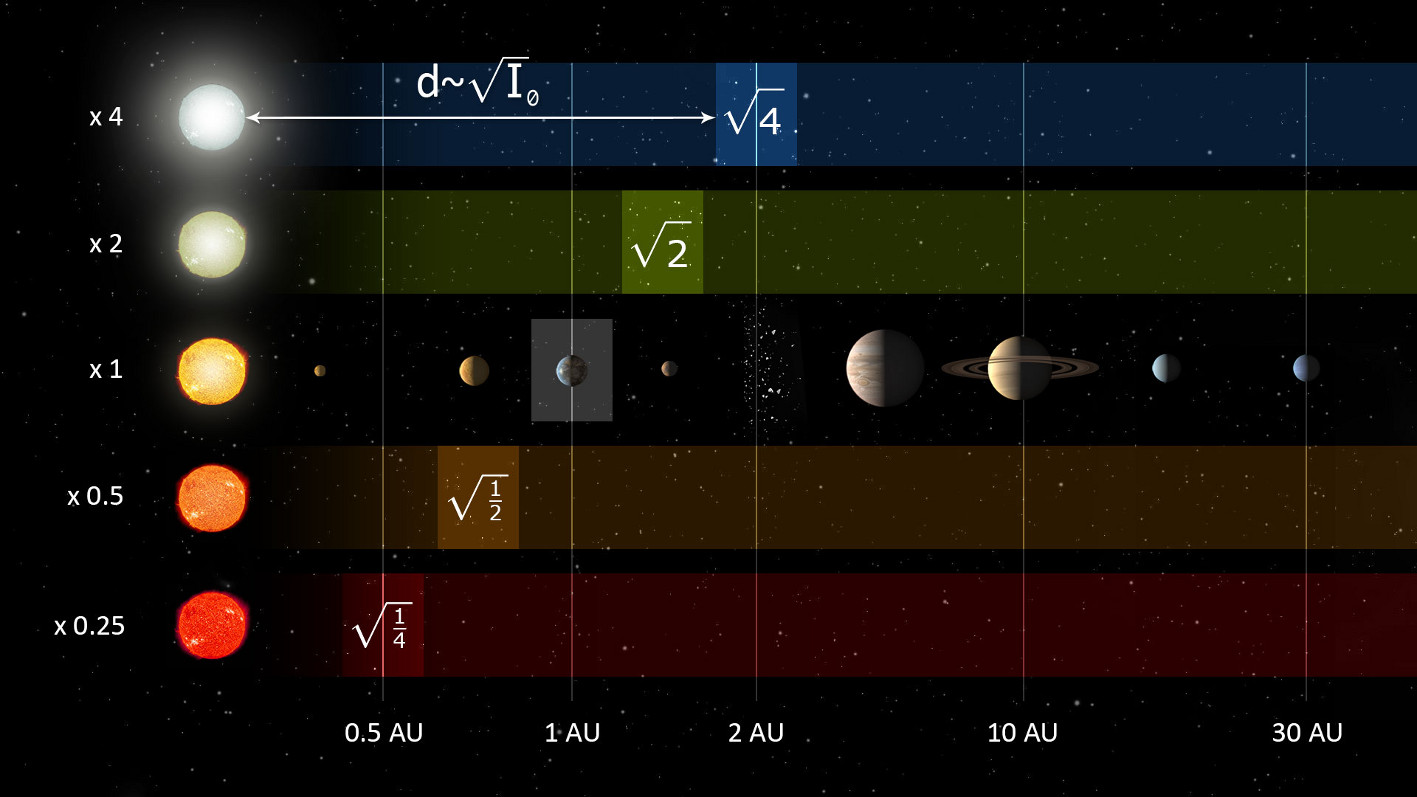}}
\caption{Distances of habitable zones for stars of varying luminosity (Derpedde, \capurl{https://de.wikipedia.org/wiki/Datei:Solarsystemau_habit.jpg}, ``Solarsystemau habit'', \capurl{https://creativecommons.org/licenses/by-sa/3.0/legalcode}).}
\label{f:hzscale}
\end{figure}

With the Earth representing the template of a habitable planet, one can calculate at which distance from any given star it must be to find comparable conditions. Of first order, this distance scales with the luminosity of the star.
\begin{equation}
d_{\rm HZ}(AU) = \sqrt{\frac{L_\star}{L_\sun}}
\end{equation}
Here, $L_\star$ is the luminosity of the star and $L_\sun$ is the solar luminosity. The distance $d_{\rm HZ}$ may then represent the location of the habitable zone (Fig.~\ref{f:hzscale}). Its width depends on what temperatures to expect at the inner and the outer edge. There are quite complex models that even include planetary properties like the atmospheric conditions and the mass. Therefore, the habitable zone is not really a well-defined range inside a planetary system but it depends on many properties that allow water to be present in its liquid form.

\subsection{Thermal radiation of solid bodies (optional information for teachers of higher term classes)}
Any given body with a temperature above absolute zero, i.e.~$T > 0\,{\rm K}$ radiates. According to Planck's Radiation Law (ideally for a black body) the distribution of frequencies of the emitted spectrum depends on the temperature.
\newglossaryentry{planck}{name={Planck's Radiation Law},description={Planck's law describes the spectral density of electromagnetic radiation emitted by a black body in thermal equilibrium at a given temperature $T$. The law is named after Max Planck, who proposed in 1900 that the electromagnetic spectrum of a black body only depends on its temperature.}}
\newglossaryentry{bb}{name={Black Body},description={A black body is an idealised physical body that absorbs all incident electromagnetic radiation, regardless of frequency or angle of incidence.}}
\newglossaryentry{spectraldens}{name={Spectral Density},description={The spectral density describes how energy is distributed across the spectrum of electromagnetic radiation.}}
\begin{equation}
U(\nu,T)=\frac{8\pi h\nu^3}{c^3}\cdot\frac{1}{\exp\left(\frac{h\nu}{kT}\right)-1}
\end{equation}
Integrating over all frequencies leads to the Radiation Law of Stefan-Boltzmann
\newglossaryentry{boltzmann}{name={Stefan-Boltzmann Law},description={The Stefan–Boltzmann law describes the power radiated from a black body in terms of its temperature. Specifically, the Stefan–Boltzmann law states that the total energy radiated per unit surface area of a black body across all wavelengths per unit time is directly proportional to the fourth power of the black body's thermodynamic temperature $T$.}} which describes the total emitted power. In reality black bodies only exist up to a certain degree of approximation. Therefore, Kirchhoff's Radiation Law
\newglossaryentry{kirchhoff}{name={Kirchhoff's Radiation Law}, description={Kirchhoff's law of thermal radiation refers to wavelength-specific radiative emission and absorption by a material body in thermodynamic equilibrium, including radiative exchange equilibrium. The emissive power of an arbitrary opaque body of fixed size and shape at a definite temperature can be described by a dimensionless ratio, called the emissivity, the ratio of the emissive power of the body to the emissive power of a black body of the same size and shape at the same fixed temperature.}} about the radiant power of any given body can be applied:
\begin{equation}
L_{\Omega,\nu}(\beta,\varphi,\nu,T)=L^0_{\Omega,\nu}(\nu,T)\cdot\alpha'_\nu(\beta,\varphi,\nu,T)
\end{equation}
This means that the radiant power $L_{\Omega,\nu}(\beta,\varphi,\nu,T)$ of a given body is as large as the total radiant power of a black body $L^0_{\Omega,\nu}(\nu,T)$ having the same temperature and the same absorptivity $\alpha'_\nu(\beta,\varphi,\nu,T)$. The spectral radiance and the absorptivity may also depend on the angle of incident radiation. Thus, according to Kirchhoff's Radiation Law, the radiant power of any given body is directly proportional to the radiant power of a black body with the same temperature. Furthermore, Kirchhoff's Radiation Law leads to the conclusion that with a given temperature a body with good heat absorption also releases heat well.

\subsection{Stefan Boltzmann's Law (optional information for teachers of higher term classes)}
The Stefan Boltzmann's Law describes the emitted power of a black body radiating isotropically into all directions. For the total radiation power of a black body the following equation applies:
\begin{equation}
P=\sigma\cdot A \cdot T^4
\end{equation}
$A$ is the area of the radiating cross section of the body and $\sigma$ is the Stefan-Boltzmann's constant, a natural constant with the value of:
\begin{equation}
\sigma = \frac{2\pi^5k^4_B}{15h^3c^2} = (5.670367 \pm 0.000013)\cdot 10^{-8}\,\frac{\rm W}{{\rm m}^2{\rm K}^4}
\end{equation}
Together with Kirchhoff's Radiation Law the Stefan Boltzmann’s Radiation Law for any given body results to
\begin{equation}
P=\epsilon(T)\cdot A \cdot T^4
\end{equation}
with a temperature dependent emissivity $\epsilon(T)$. The emissivity is a parameter that reflects the properties of the radiating substance and surface. For a given temperature, surfaces with different emissivity appear differently bright. As a result, the thermal radiation is also stronger.

\begin{figure}
\centering
\resizebox{\hsize}{!}{\includegraphics{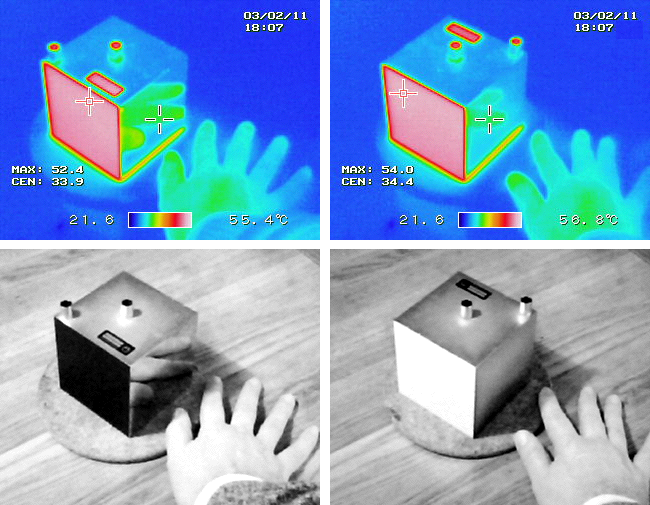}}
\caption{Photographs of Leslie's cube. The colour photographs are taken using an infrared camera; the black and white photographs underneath are taken with an ordinary camera. All faces of the cube are at the same temperature of about $55\degr$C. The face of the cube that has been painted has a large emissivity, which is indicated by the reddish colour in the infrared photograph. The polished face of the aluminium cube has a low emissivity indicated by the blue colour, and the reflected image of the warm hand is clear (Pieter Kuiper, \capurl{https://commons.wikimedia.org/wiki/File:LesliesCube.png}, public domain).}
\label{f:cube}
\end{figure}

\subsection{Radiation balance (optional information for teachers of higher term classes)}
A body that absorbs energy in form of radiation warms up as long as the energy emitted is as high as the energy that is absorbed. This means that body is in thermal equilibrium. This scenario can applied to the Earth that receives electromagnetic radiation of the Sun with a flux density (solar constant) of:
\newglossaryentry{E}{name={Solar Constant},description={It describes the value of the flux density -- the radiative power per unit area -- for solar radiation at the Earth's orbit.}}
\begin{equation}
E_0=1367\frac{\rm W}{{\rm m}^2}
\end{equation}
If the Earth is in thermal equilibrium, the energy emitted must be as high as the energy absorbed. It is important that only half of the Earth is irradiated at the same time. As part of the sunlight hits the surface in an angle not perpendicular, only an area as big as the cross section of the Earth is illuminated. Thus, the energy absorbed by the Earth is:
\begin{equation}
P=E_0\cdot\pi\cdot r_e^2 \approx 1.74\cdot 10^17\,{\rm W}
\end{equation}
Here $r_e$ is the Earth's radius. The energy emitted by the Earth can be calculated with Stefan-Boltzmann’s law. Here, it is important that the complete surface radiates.
\begin{equation}
P=\sigma \cdot A \cdot T^4
\end{equation}
With this, the temperature can be calculated, if the Earth did not have an atmosphere.
\begin{equation}
T=\sqrt[4]{\frac{P}{\sigma\cdot A}}=\sqrt[4]{\frac{P}{\sigma\cdot 4\cdot\pi\cdot r_e^2}}\approx 279\,{\rm K}
\end{equation}

\subsection{A model for circumstellar habitable zone (optional information for teachers of higher term classes)}
In the following, a simple model for characterising the circumstellar habitable zone has to be found. It is purely based on the assumption of thermal equilibrium. The total radiative power emitted by the Sun (luminosity) is derived from the solar constant, the surface of a sphere and the distance between the Sun and the Earth. Since the sun radiates in all directions isotropically, we can assume that the solar constant is valid for any point on such a sphere. For $r^2_{se}$ representing the distance between the Sun and the Earth find:
\begin{equation}
L=P_s=E_0\cdot 4\pi r_{se}^2\approx 3.845\cdot 10^{26}\,{\rm W}
\end{equation}
The solar constant for any given distance r is:
\begin{equation}
E_0(r)=\frac{L}{4\pi r^2}=E_0\cdot \frac{r^2_{se}}{r^2}
\end{equation}
This demonstrates that -- as expected -- the received flux density of a constant radiation source depends on the inverse square of the distance. Following the applied logic of the thermal equilibrium the effective temperature of a planet with the radius $r_ p$ results to:
\begin{eqnarray}
E_0(r) \cdot\pi\cdot r_p^2 &=& \sigma\cdot A \cdot T^4 \\
                           &=& \sigma \cdot 4\pi r_p^2 \cdot T^4\\
       \Leftrightarrow T^4 &=& \frac{E_0(r)}{4\sigma}=\frac{E_0\cdot\frac{r^2_{se}}{r^2}}{4\sigma}\\
             \Rightarrow T &=& \sqrt[4]{\frac{E_0\cdot r^2_{se}}{4\sigma r^2}}=\frac{\alpha}{\sqrt{r}}
\end{eqnarray}
With the constant
\begin{equation}
\alpha = \sqrt[4]{\frac{E_0\cdot r^2_{se}}{4\sigma}}
\end{equation}
which for the Sun yields:
\begin{equation}
\alpha = 1.08\cdot 10^8\frac{{\rm m}^{0.5}}{\rm K} = 278.58\frac{{\rm AU}^{0.5}}{\rm K}
\end{equation}
By selecting realistic limitations for the average temperature of a planet, the limits for a circumstellar habitable zone can be determined.

\section{List of material}
Since photovoltaic cells tend to be quite expensive, teachers may prepare a restricted number of experimental set-ups. The activities need the following items.

\subsection{Activity 1}
\begin{itemize}
 \item Strong lamp, floodlight
 \item Dimmer to regulate the brightness
 \item Folding rule or yardstick
 \item Photovoltaic cell with attached electric motor or fan (be sure that the power the cell generates is big enough to drive the motor)
\end{itemize}

\subsection{Activity 2}
\begin{itemize}
 \item  Pencils (regular and coloured)
 \item Compass (drawing tool)
 \item Millimetre paper
 \item Ruler
 \item Calculator
 \item Computer with internet access (can be only one operated by the teacher)
\end{itemize}

\section{Target group details}

\noindent
Suggested age range: 12 -- 16 years\\
Suggested school level: Middle School, Secondary School\\
Duration: 1 hour

\section{Goals}
With this activity the students

\begin{itemize}
\item  understand the qualitative correlation between the distance of a planet to a star and the energy density that impacts the planet.
 \item learn that liquid water in a planetary system can only be present in a confined corridor, the habitable zone.
 \item understand that searching for extra-terrestrial life as we know it means to look for conditions to keep water liquid.
 \item understand that the size and extent of a habitable zone depends on the luminosity of a star.
\end{itemize}

\section{Learning objectives}
\begin{itemize}
\item The students will be able to explain what are the key conditions needed to find potential life outside Earth.
\item The students will be able to explain what the habitable zones of planetary systems and their characteristics are.
\item The students will be able to demonstrate how the flux density received by a body depends on its distance from the source.
\end{itemize}

\section{Evaluation}
The students should be able to explain the phenomenon observed in their own wording. They should also be able to record and analyse data on their own and draw the necessary conclusions for their environment. With support they should be able to explain their observations correctly. In case the necessary physical basic concepts have not been reviewed yet the students should be able to explain their observations in their own wording.

Ask the students if there is life on Mars and if there is water on Mars. Ask the students if there is any relation between them.

Ask the students what it takes to support life on other planets.

Ask the students what conditions are needed to keep water liquid.

Ask students what provides us and other planets with the heat to keep water liquid.

\section{Full description of the activity}
\subsection{Introduction}
A key component of sustaining life on a planet is its capability of providing liquid water. For this, it is important that the receiving radiative power of the central star is big enough to raise the temperatures on the planet above a value that at least partially permits water to be liquid. This constellation can be seen as an engine of life that needs enough power to start.

Introduce the topic by asking about the solar system (planets, composition) and where in that system life had evolved. Perhaps, spark a discussion about the conditions of sustaining life on planets different than Earth (e.g. Venus, Mars). Try to direct the discussion towards the presence of water.

\medskip\noindent
Q: Is there life on Mars?\\\noindent
A: We haven’t found anything that would point towards that direction.

\medskip\noindent
Q: Is there liquid water on Mars?\\\noindent
A: Very little to none. At least not so much that we can expect sustained open water areas or running rivers.

\medskip\noindent
Q: Why are those two things related?\\\noindent
A: Water is the key to life. Without water, there is no life the way we know it.

\medskip\noindent
Ask the students, if they have heard about planets around stars other than the sun.

\medskip\noindent
Q: What does it take to support life on other planets?\\\noindent
A: There are a few conditions that help sparking and sustaining life, but the key is – again – water. Heat is only needed to keep the water liquid, at least in some places. Other boundary conditions that may help are energy sources (light, chemical) or magnetic fields to fend off ionising particle radiation. But as long life evolves in oceans, this does not really matter.

\medskip\noindent
Q: What conditions are needed to keep water liquid?\\\noindent
A: Heat, and perhaps salt. Salt helps to permit lower temperatures to keep water liquid.

\medskip\noindent
Q: What provides us and other planets with the heat needed to keep water liquid?\\\noindent
A: Stars, the Sun. A greenhouse effect provided by an atmosphere helps to raise temperatures.

\medskip\noindent
Q: What happens to water, if it is very cold or very hot?\\\noindent
A: It freezes or boils and evaporates.

\medskip\noindent
The following experiment is a simple analogue, in which the star is represented by a lamp, and the
photovoltaic cell combined with the motor stands for the planet.

\subsection{Activity 1: Engine of Life}
\subsubsection{Materials needed}
\begin{itemize}
\item  Strong lamp, floodlight
\item Dimmer to regulate the brightness
\item Folding rule or yardstick
\item Photovoltaic cell with attached electric motor or fan (be sure that the power the cell generates is big enough to drive the motor)
\end{itemize}

\subsubsection{Experimental set-up}
\begin{enumerate}
\item Solder the motor to the photovoltaic cell.
\item A coloured disc made of cardboard attached to the rotation axis improves the observation. If a fan is used, the wings may be painted.
\item Plug the lamp into the dimmer and then the dimmer into the socket.
\end{enumerate}

\begin{figure}
\centering
\resizebox{\hsize}{!}{\includegraphics{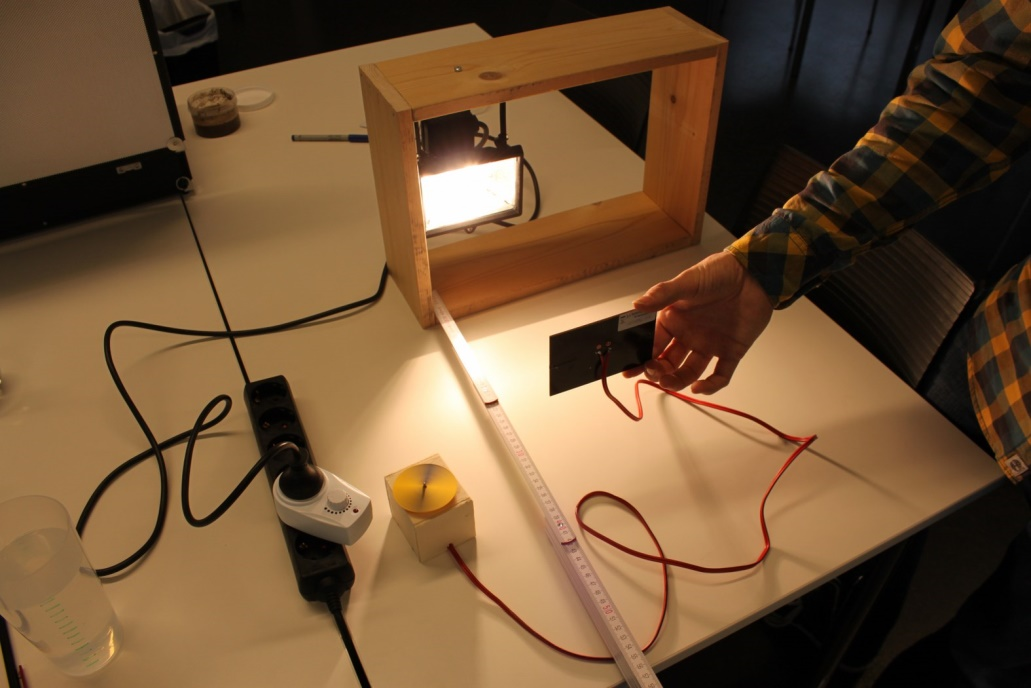}}
\caption{Experimental set-up (own work).}
\label{f:a1}
\end{figure}

\subsubsection{Experimental procedure}
\begin{enumerate}
\item Switch on the lamp.
\item Hold the photovoltaic cell at a large distance from the lamp. The motor should not move.
\item Approach the lamp and measure the distance at which the electric motor starts moving.
\item Repeat this procedure for different brightness settings of the lamp by using the dimmer.
\end{enumerate}

\begin{figure}
\centering
\resizebox{\hsize}{!}{\includegraphics{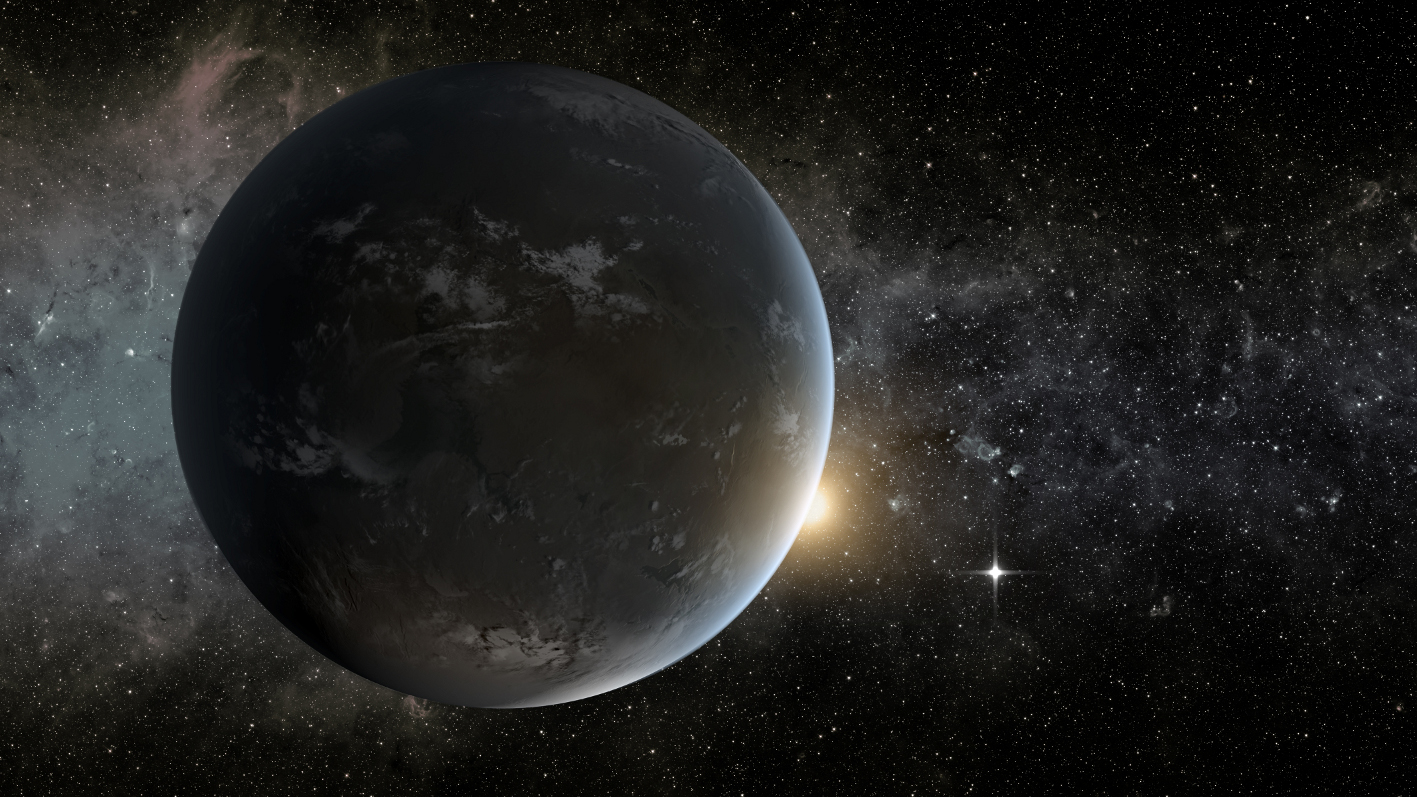}}
\caption{Artistic impression of the Kepler-62f exoplanet (NASA Ames/JPL-Caltech).}
\label{f:kepler62}
\end{figure}

\subsubsection{Common tasks}
\begin{enumerate}
\item Write down your observations. Describe the results you get when varying the brightness of
the lamp.\\\noindent
Expected result: The distance necessary for the electric motor to begin to move is smaller
with a dimmed lamp than with a bright lamp.
\item Transferring this model experiment to the configuration of the solar system, life sustaining
conditions (running motor) are possible, because the Earth (photovoltaic cell) is close
enough to the Sun (lamp). It is the outer edge of the habitable zone. What does the
experiment tell us about exoplanets in other planetary systems with different stars that are
supposed to harbour life?\\\noindent
Expected result: A planet in a planetary system with a star that less bright than the Sun
needs to be located closer to then sun in order to be in the habitable zone.
\end{enumerate}

\subsubsection{Tasks for ages 14 and higher}
\begin{enumerate}
\item What happens to the motor, when the cell is very close to the lamp?\\\noindent
Expected result: It runs very fast. It ``overheats''.
\item Can we expect planets sustaining life to be at any distance inside the inner edge of the
habitable zone?\\\noindent
Expected result: No, because with decreasing distance the receiving power rises. This leads
to an increased heating of the planet. If the temperature is too high, water cannot remain
liquid.
\end{enumerate}

\subsection{Activity 2: The Habitable Zone of Kepler-62}
\subsubsection{Materials needed}
\begin{itemize}
\item Pencils (regular and coloured)
\item Compass (drawing tool)
\item Millimetre paper
\item Ruler
\item Calculator
\item Computer with internet access (can be only one operated by the teacher)
\end{itemize}

Kepler-62 is a star which is a little cooler and smaller than the Sun. It is part of the constellation Lyra. In 2014, it was discovered by the Kepler Space Telescope (see: \url{https://www.nasa.gov/mission_pages/kepler/overview/index.html}) as a star that hosts five planets orbiting it. The details about the star Kepler-62 are summarised in Tab.~\ref{t:kepler62}.

\begin{figure}
\centering
\resizebox{\hsize}{!}{\includegraphics{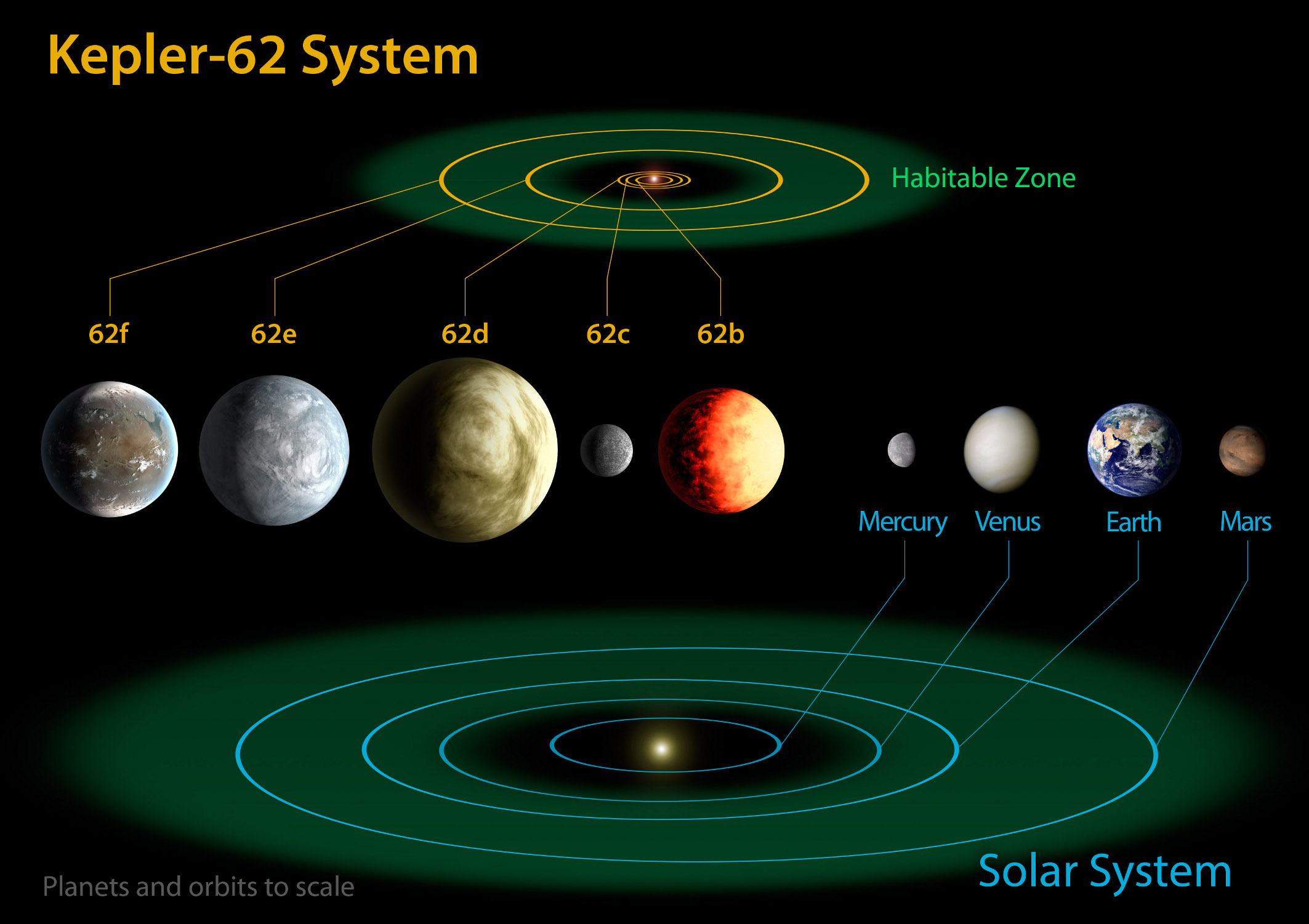}}
\caption{ Comparison between the solar and Kepler-62 planetary systems (NASA Ames/JPL-Caltech).}
\label{f:kepler62system}
\end{figure}

\begin{table}[!ht]
\centering
\caption{Properties of the star Kepler-62.}
\label{t:kepler62}
\begin{tabular}{ll}
\hline\hline
Property      & Value \\
\hline
Name          & Kepler-62 \\
Distance      & ca. 368~pc \\
Spectral Type & K2V \\
Luminosity    & $0.21\,L_\sun$ \\
Radius        & $0.64\,R_\sun$ \\
Mass          & $0.69\,M_\sun$ \\
Surface temperature & 4925\,K \\
\hline
\end{tabular}
\end{table}

Some of the five exoplanets are suspected to be Earth-like. The main properties of the five planets are provided in Tab.~\ref{t:kepler62sys}. All planetary orbits are nearly circular and listed in Astronomical Units (AU). This is the mean distance between the Sun and the Earth.

\begin{table}[!ht]
\centering
\caption{Properties of the five exoplanets in the Kepler-62 system.}
\label{t:kepler62sys}
\begin{tabular}{ccc}
\hline\hline
Name        & Orbital radius (AU) & Mass (Earth masses) \\
\hline
Kepler-62 b & 0.0553 & ca. 2.1 \\
Kepler-62 c & 0.093  & ca. 0.1 \\
Kepler-62 d & 0.120  & ca. 5.5 \\
Kepler-62 e & 0.427  & ca. 4.5 \\
Kepler-62 f & 0.712  & ca. 2.8 \\
\hline
\end{tabular}
\end{table}

\subsubsection{Tasks (Drawing a scaled model)}
Determine or discuss a suitable scale that allows you to put the entire system on a sheet of paper. Fill the Tab.~\ref{t:kepler62scaled} with the scaled orbital radii. Round up the values to full millimetres.

The model will show the planetary system from a bird's eye view. Use the compass to draw the scaled circular orbits around the assumed position of the host star Kepler-62. In the next step, you will add the habitable zone. First, you can apply the simple equation
\begin{equation}
d_{\rm HZ}(AU) = \sqrt{\frac{L_\star}{L_\sun}}
\end{equation}
that only depends on the luminosity of the star. Note that this equation tells you, where an Earth-like planet would be located around a Sun-like star of lower luminosity. Use Tab.~\ref{t:kepler62} to calculate the distance of the habitable zone from Kepler-62 and add it to Tab.~\ref{t:kepler62scaled}.

\begin{figure}
\centering
\resizebox{\hsize}{!}{\includegraphics{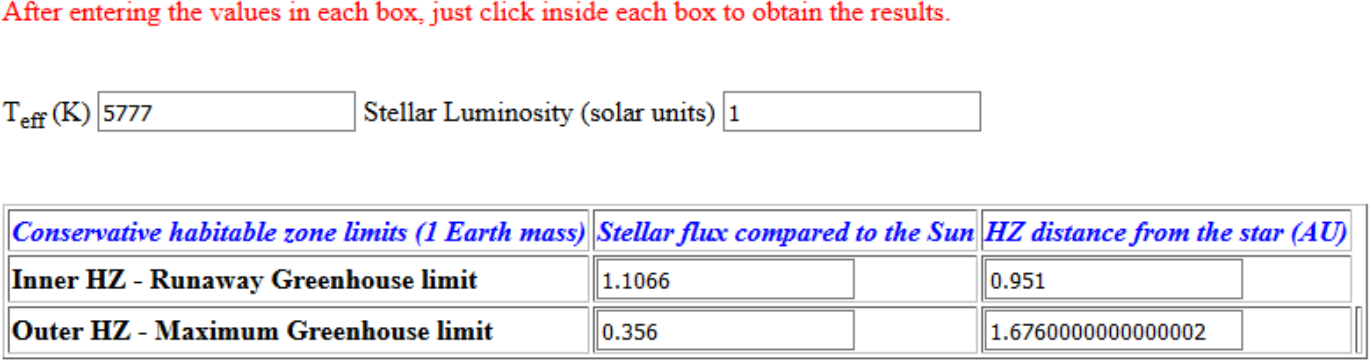}}
\caption{Screen shot of the on-line tool to calculate the dimensions of a habitable zone. The example above show the values of the Sun. Important: After inserting the values, you have to click into the field to submit the entry (\capurl{http://depts.washington.edu/naivpl/sites/default/files/hz.shtml}).}
\label{f:hzcalc}
\end{figure}

Calculating the proper boundaries of habitable zones can be tricky and need sophisticated models. There is an on-line tool at
\url{http://depts.washington.edu/naivpl/sites/default/files/hz.shtml} that does the calculations, when the luminosity and the surface temperature of the star is provided (This can be done by the teacher instead of handing out computers for all the students.). All you have to do is enter the surface temperature $T_{\rm eff}$ and the luminosity of the star in the upper two fields as shown in Fig.~\ref{f:hzcalc}. Note that after entering a value, you have to click into the field again. Use the values of Kepler-62 from Tab.~\ref{t:kepler62}.

Use the result labelled as ``Conservative habitable zone limits (1 Earth mass)'' and ``HZ distance from the star (AU)''. Add the distances for the inner and the outer edge of the habitable zone to the table below and calculate the scaled radii.

\medskip\noindent
Q: How well does the simple distance of the habitable zone agree with the model calculation?\\\noindent
A: The simple value is close to the inner edge of the more accurately calculated habitable zone.

\medskip\noindent
Q: Can you think of a reason, why the simple solution is so close to an extreme value of the modelled
solution? In which way do the two approaches differ? (Hint: If the students have difficulties with an
answer, let them use the on-line calculator with the solar surface temperature but the stellar
luminosity.)\\\noindent
A: The simple approach does not take into account the surface temperature.

\medskip\noindent
Q (perhaps better suited for higher term classes): How does the missing parameter (the
temperature) influence the radiative characteristics of the star Kepler-62?\\\noindent
A: Lower temperatures mean redder spectra. The spectrum of Kepler-62 contains a relatively higher
amount of infrared radiation than the Sun. This is more effective in directly heating planetary
atmospheres.

\medskip
Now add the inner and outer edge of the habitable zone to the scaled model of the planetary
system. You can paint the area between the inner and outer edges of the habitable zone. Green would be an
appropriate colour.

\medskip\noindent
Q: Which of the planets are inside the habitable zone?\\\noindent
A: Kepler-62f

\begin{table}[!ht]
\centering
\caption{Orbital parameters of the Kepler-62 planetary system (The scaled values are optimised for a sheet of A4 paper
with a width of 18~cm. They are not provided with the worksheets).}
\label{t:kepler62scaled}
\begin{tabular}{ccc}
\hline\hline
Name        & Orbital      & Scaled      \\
            & radius (AU)  & radius (cm) \\   
\hline
Kepler-62 b & 0.0553 & 0.60 \\
Kepler-62 c & 0.093  & 1.01 \\
Kepler-62 d & 0.120  & 1.30 \\
Kepler-62 e & 0.427  & 4.64 \\
Kepler-62 f & 0.712  & 7.74 \\
\hline
Habitable Zone (simple) & 0.458 & 4.98 \\
Habitable Zone (inner) & 0.456 & 4.96 \\
Habitable Zone (outer) & 0.828 & 9.00 \\
\hline
\end{tabular}
\end{table}

\begin{figure}
\centering
\resizebox{0.6\hsize}{!}{\includegraphics{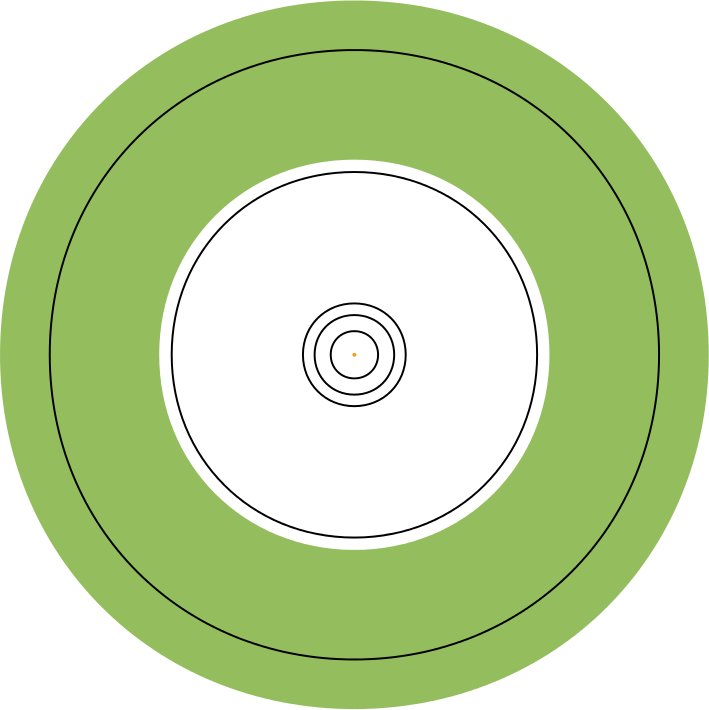}}
\caption{Model of the Kepler-62 system (solution not presented to the students).}
\label{f:model}
\end{figure}

\section{Connection to school curriculum}
This activity is part of the Space Awareness category ``Our Fragile Planet'' and related to the curricula topics:

\begin{itemize}
\item Composition and structure
\item Orbit and rotation
\item Habitability
\end{itemize}

\section{Conclusion}
The activity teaches about habitable zones of planetary systems. It comprises a simple experiment
that demonstrates how the flux density received by a body depends on its distance from the source.
It consists of a strong lamp, a photovoltaic cell that receives its power from the lamp and drives a
motor as soon as the power is sufficiently high. This is relevant for understanding the dimensions of
a habitable zone, where liquid water can be present for certain boundary conditions. The running
motor in the experiment depicts the running engine of life within a planetary system. In addition, a
second activity helps the students understand the composition of a real exoplanetary system and its
habitable zone by calculating and drawing a scaled model of it.

\begin{acknowledgements}
This resource was developed in the framework of Space Awareness. Space Awareness is funded by the European Commission’s Horizon 2020 Programme under grant agreement no. 638653.
\end{acknowledgements}

\bibliographystyle{aa}
\bibliography{FragilePlanet}

\begin{thebibliography}{5}
\expandafter\ifx\csname natexlab\endcsname\relax\def\natexlab#1{#1}\fi

\bibitem[{{Anglada-Escud{\'e}} {et~al.}(2014){Anglada-Escud{\'e}}, {Arriagada},
  {Tuomi}, {Zechmeister}, {Jenkins}, {Ofir}, {Dreizler}, {Gerlach}, {Marvin},
  {Reiners}, {Jeffers}, {Butler}, {Vogt}, {Amado},
  {Rodr{\'{\i}}guez-L{\'o}pez}, {Berdi{\~n}as}, {Morin}, {Crane}, {Shectman},
  {Thompson}, {D{\'{\i}}az}, {Rivera}, {Sarmiento}, \&
  {Jones}}]{2014MNRAS.443L..89A}
{Anglada-Escud{\'e}}, G., {Arriagada}, P., {Tuomi}, M., {et~al.} 2014, \mnras,
  443, L89

\bibitem[{{European Space Agency}(2016)}]{exomars}
{European Space Agency}. 2016, Robotic Exploration of Mars,
  \href{http://exploration.esa.int/mars/46048-programme-overview/}{http://exploration.esa.int/mars/46048-programme-overview/}

\bibitem[{{Mayor} \& {Queloz}(1995)}]{1995Natur.378..355M}
{Mayor}, M. \& {Queloz}, D. 1995, \nat, 378, 355

\bibitem[{Ratner(2017)}]{ratner_2017}
Ratner, P. 2017, NASA Discovers Why Saturn's Moon Enceladus May Be the Best
  Place to Look for Alien Life, Bigthink.com,
  \href{http://bigthink.com/paul-ratner/nasa-discovers-why-saturns-moon-enceladus-may-be-the-best-place-to-look-for-alien-life}{http://bigthink.com/paul-ratner/nasa-discovers-why-saturns-moon-enceladus-may-be-the-best-place-to-look-for-alien-life}

\bibitem[{Redd(2015)}]{redd_2015}
Redd, N.~T. 2015, An Ocean Flows Under Saturn's Icy Moon Enceladus, Space.com,
  \href{https://www.space.com/30559-saturn-moon-enceladus-has-ocean.html}{https://www.space.com/30559-saturn-moon-enceladus-has-ocean.html}

\end{thebibliography}

\glsaddall
\printglossaries

\begin{appendix}
\section{Relation to other educational materials}
This unit is part of a larger educational package called “Our Fragile Planet” that introduces several historical and modern techniques used for navigation. An overview is provided via: \href{http://www.space-awareness.org/media/activities/attach/6782cf82-139d-4e69-b321-f77312be1fcc/The\%20Climate\%20Box\%20an\%20educational\%20ki\_2udsFBb.pdf}{Our\_Fragile\_Planet.pdf}

\section{Supplemental material}
The supplemental material is available on-line via the Space Awareness project website at \url{http://www.space-awareness.org}. The direct download links are listed as follows:
 
\begin{itemize}
\item Worksheets: \href{https://drive.google.com/file/d/0Bzo1-KZyHftXZDBMQU1kdjVlaFE/view?usp=sharing}{astroedu1624-The\_Engine\_of\_Life-WS.pdf}
\end{itemize}

\end{appendix}
\end{document}